\documentclass[conference]{article}

\usepackage{cite}
\usepackage{amsmath,amsfonts,amssymb}
\usepackage{graphicx}
\usepackage{float}
\usepackage{subfig}
\usepackage{authblk}
\usepackage[colorlinks=true, allcolors=blue]{hyperref}
\usepackage[a4paper,top=3cm,bottom=2cm,left=3cm,right=3cm,marginparwidth=1.75cm]{geometry}

\providecommand{\keywords}[1]
{
  \small	
  \textbf{\textit{Keywords---}} #1
}

\begin{document} 

\title{Smart-readout of the Skipper-CCD: Achieving Sub-electron Noise Levels in Regions of Interest}

\author[1]{Fernando Chierchie}
\author[2]{Guillermo Fernandez Moroni}
\author[2]{Leandro Stefanazzi}
\author[3]{Claudio Chavez}
\author[1,6]{Eduardo Paolini}
\author[2]{Gustavo Cancelo}
\author[4]{Miguel Sofo Haro}
\author[2]{Javier Tiffenberg}
\author[2]{Juan Estrada}
\author[5]{Sho Uemura}

\affil[1]{Instituto de Investigaciones en Ingenier\'{i}a El\'{e}ctrica ``Alfredo C. Desages'' (IIIE-CONICET)\\
Departamento de Ingenier\'{i}a El\'{e}ctrica y de Computadoras, Universidad Nacional del Sur (UNS), Bah\'ia Blanca, Argentina.}
\affil[2]{Fermi National Accelerator Laboratory, Batavia IL, United States.}
\affil[3]{Facultad de Ingenier\'{i}a - Universidad Nacional de Asunci\'{o}n, Paraguay, Asunci\'{o}n, Paraguay.}
\affil[4]{Centro At\'omico Bariloche and Instituto Balseiro, Comisi\'on Nacional de Energ\'ia At\'omica (CNEA), Universidad Nacional de Cuyo (UNCUYO), Rio Negro, Argentina.}
\affil[5]{School of Physics and Astronomy,Tel-Aviv University, Tel-Aviv, Israel.}
\affil[6]{Comisi\'on de Investigaciones Cientif\'icas Prov. Buenos Aires (CICpBA), Argentina.}

\maketitle

\begin{abstract}
The skipper CCD is a special type of charge coupled device in which the readout noise can be reduced to sub-electron levels by averaging independent measurements of the same charge. Thus the charge in the pixels can be determined by counting the exact number of electrons. The total readout time is proportional to the number of measurements of the charge in each pixel. For some applications this time may be too long; however,  researchers usually are interested only on certain region within the matrix of pixels. In this paper we present the development of a smart skipper readout technique that allows the user to specify regions of interest of the CCD matrix where an arbitrary (high) number of measurements of the same charge can taken to obtain the desired noise level, and far less measurements are performed in those regions that are less interesting to the researcher, therefore reducing the total readout time.
\end{abstract}


\section{Introduction}
\label{sec:intro}  

The Skipper-CCDs, is a special type of charge-coupled device (CCD) sensor that can achieve sub-electron readout noise levels \cite{skipper_2012,Tiffenberg:2017aac}. With the recent development of the readout electronics specifically designed for the skipper \cite{haro2017low, LTA8709274,cancelo2020low} many scientific experiments are being developed using this sensor. Some of them are in the fields of particle physics such as light dark matter searches \cite{Barak2020, OSCURA2020, DAMIC2019} and reactor neutrino detection experiments \cite{violeta2020, CONNIE2019}; other experiments are for high precision silicon property measurements. In \cite{Rodrigues_2020} an absolute measurement of the variance and the mean distribution (fano factor)  of the charge produced by an X-rays source in Silicon at $5.9$ keV was presented and charge collection efficiency is studied in \cite{fernandezmoroni2020charge}.
Other fields such as quantum imaging are also possible to explore with the skipper CCD \cite{Kuk_2020}.

The sub-electron noise in the skipper CCD is achieved by making independent measurements of the same pixel charge and averaging the result. The noise is reduced by the square root of the number of samples $N$ at the expense of an increase of the readout time proportional to $N$. Some applications would benefit from a reduced readout time while keeping the sub-electron noise. In this paper we present the results of a smart readout technique of the Skipper-CCD based on regions of interest, in which each pixel of these regions can be read $N$ times (to achieve a desired noise level), and  performing less readouts in other regions to speed up the readout process at the expense of higher noise in those non-interest zones. 

In Section \ref{Sec:ReviwePixel} we review the digital computation of the pixel value for standard and Skipper-CCD. In Section \ref{Sec:ROI_Basline} we introduce the proposed smart skipper readout technique and present the problem of baseline variation and compensation. Finally, in Section \ref{sec:Exp} experimental results with measured images using the proposed approach are presented and conclusions are drawn in Section \ref{sec:conclu}.

\section{Review of CCD video signal and pixel computation}
\label{Sec:ReviwePixel}

\begin{figure}
    \centering
    \includegraphics[width=0.95\textwidth]{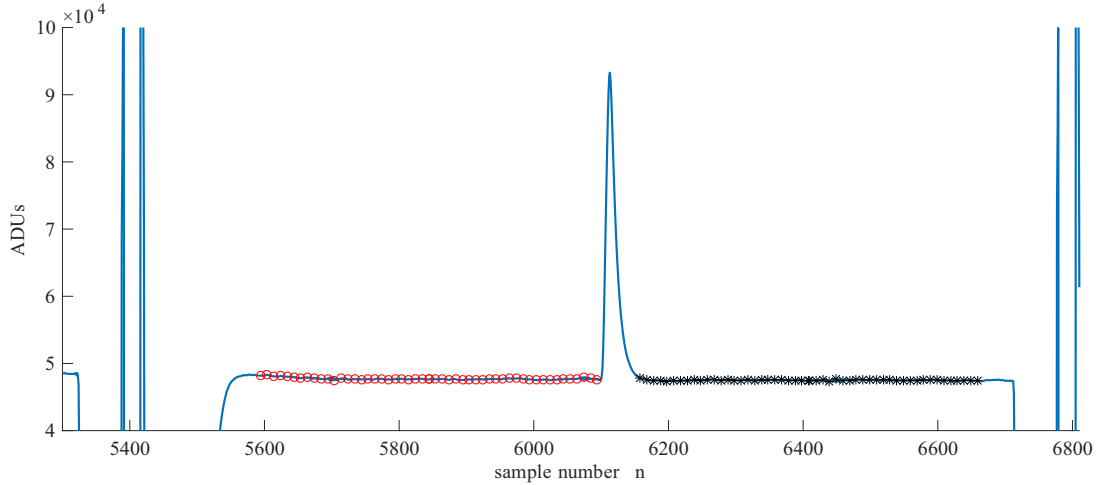}
    \caption{One empty pixel of the video signal. The ``$\circ$'' indicates the $M$ samples of the pedestal while the ``$*$'' the $M$ samples of the signal level. The peak between the two intervals is caused by clock feedthrough when the charge is transferred to sense node: $n_0$ samples are discarded to avoid these samples.}
    \label{fig:VideoSamples}
\end{figure}

The video signal of the CCD has two main intervals: the reference or pedestal interval and the signal interval. In the pedestal level the sense node is reseted to a reference voltage while in the signal level the charge is dumped into the sense node. Figure \ref{fig:VideoSamples} shows a measurement of one pixel including the pedestal and signal samples. The value of each pixel is usually computed using the dual slope integration (DSI) method \cite{janesick2001scientific}. In the digital implementation of the DSI, integrals are replaced by sampling averaging and the pixel value is given by
\begin{equation}
P_i = \frac{A}{M} \left( \sum_{k=M+n_0}^{2M+n_0} x[n] - \sum_{k=0}^{M} x[n] \right),
\label{eq:Pi}
\end{equation}
where $x[n]=x(t)|_{t=nT_s}$ is the sampled video signal, $M$ is the number of averaged samples in the pedestal and signal intervals, $n_0$ is the number of samples discarded between the pedestal and signal levels, the interval $[0,M]$ is the pedestal interval, the interval $[M+n_0, 2M+n_0]$ is the signal interval and $A$ is an arbitrary gain of the signal processing chain.
We define the integration time as $t_i=M T_s$.

The Skipper-CCD has floating gate sensing node that allows multiple non destructive measurements of the same pixel charge to be averaged to reduce the noise up to sub-electron levels. In this case, the final pixel value is 
\begin{equation}
P_{i,skp} = \frac{1}{N}\sum^{N-1}_{j=0} P_i[j]
\label{eq:pixel with skipper}
\end{equation}
where, for each value of $j$, $P_i[j]$ represent a new measurement of pixel-$i$ computed with (\ref{eq:Pi}). This averaging reduces the standard deviation of the readout noise to  \cite{Tiffenberg:2017aac,cancelo2020low}
\begin{equation}
\sigma_{P_{i,skp}}=\sigma_{P_i}/\sqrt{N}.
\label{Ec:ReduccionRuido}
\end{equation}

\section{Region of interest with variable number of skipper samples}
\label{Sec:ROI_Basline}
\begin{figure}
    \centering
    \includegraphics[width=0.95\textwidth]{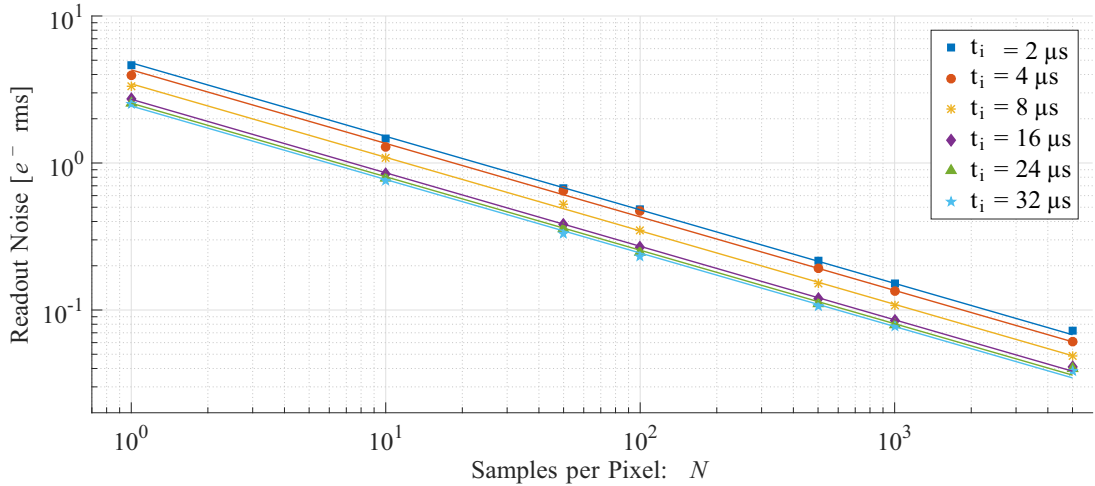}
    \label{Fig:skipperScanParamSsamp}
    \caption{Skipper scan measurements: sweep of the readout noise as a function of the number of samples $N$ taken for each pixel parametrized for different integration times: $t_i=2, 4, 8, 16, 24, 32$ $\mu s$. Measurements are indicated with markers and the straight lines show the theoretical prediction $\propto 1/\sqrt{N}$.}
    \label{Fig:skipperScan}
\end{figure}

As described by (\ref{Ec:ReduccionRuido}), the noise of a skipper CCD can be reduced by taking $N$ measurements of the same pixel charge and averaging the result as indicated in (\ref{eq:pixel with skipper}). Since the noise is independent in each measurements the variance of the pixel values reduces with the square root of $N$. This was verified experimentally \cite{cancelo2020low} as shown in Fig~\ref{Fig:skipperScan}, where both measurements and the theoretical reduction $1/\sqrt{N}$ are depicted, parametrized for different integration times $t_i$. In any case for a fixed $t_i$ the value $N$ of skipper samples per pixel  can be chosen to achieve sub-electron noise. 

The total readout time of the CCD sensor is proportional to $t_i\times N\times N_{pix}$, where $N_{pix}$ is the total number of pixels in the sensor. For certain applications the time required for reading the full skipper array with sub-electron noise might be too long. However, in some of these cases, not all the pixels in the  sensor have to be read with the same noise level (same $N$); a higher noise might be tolerated in certain regions while extremely low noise might me required in others. This results in the specification of a \textit{region of interest} (ROI) within the array of pixels in which a pre-specified noise level is expected.

% Region de interes: flexibilidad en la especificacion y  secuenciamiento de los relojes: se necesita un sequencer diferente ya no son 3 ciclos for necesito definicion pixel a pixel. Implementacion en LTA.
From a  technological perspective in terms of hardware, software, firmware and signal processing, this \textit{smart skipper} readout requires several modification to the standard CCD readout system. Most of the changes are in the sequencer, which is the firmware block in charge of generating the signals but also in the post-processing of the images. 

The signals to drive the CCD and acquire the image are usually implemented by a sequential machine in which the user specifies three nested loops: the first for the vertical clocking, the second for the horizontal clocking and the third for the $N$ skipper samples of each pixel. In smart skipper, arbitrary pixels of the CCD are readout with different precision (different $N$) and the sequencer is inherently more complex.  To implement such a system the standard (fixed) state machine was replaced with a dedicated microprocessor instantiated in the FPGA. The processor is in charge of implementing a more flexible state machine that can change the $N$ value for individual pixels or for arbitrary sized rectangular pixel regions. The user can specify these ROIs depending on the application. Further details of the architecture are discussed in Section \ref{sec:Exp}.

\begin{figure}
    \centering
    \includegraphics[width=0.95\textwidth]{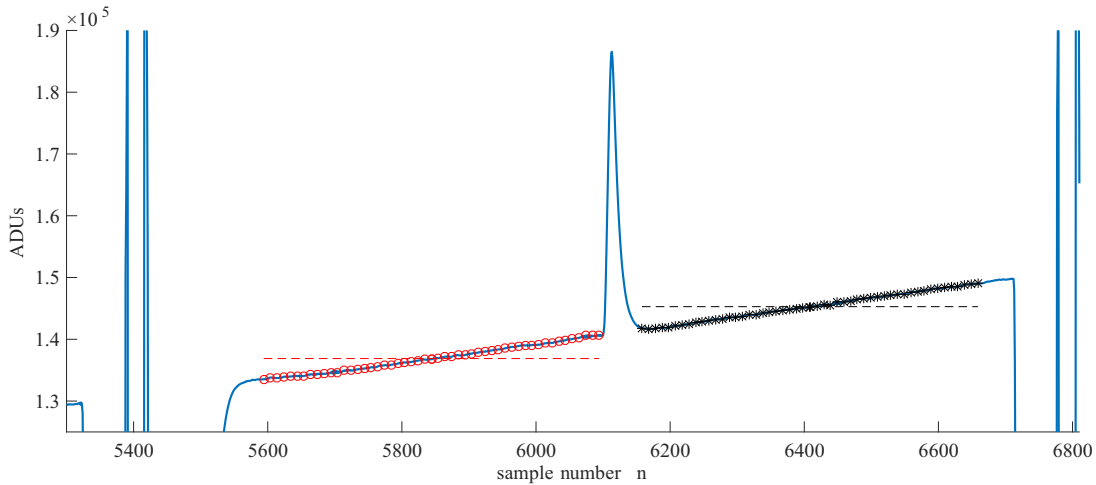}
    \caption{One empty pixel of the video signal as the one shown in Fig.~\ref{fig:VideoSamples} but affected by a transient response caused by constant changes of the clocking signal. The ``$\circ$'' indicates the $M$ samples of the pedestal and the red dashed line the average of those samples. The ``$*$'' are the $M$ samples of the signal level and the dashed black line the average.  When averaging the pedestal and signal levels and computing $P_i$ in (\ref{eq:Pi}) a nonzero value is obtained: the baseline of the pixel $BL_{i}$.}
    \label{fig:VideoSamples_BL}
\end{figure}

\subsection{Baseline}
\label{Sec:Baseline}
%Curva de baseline que se mueve e imagen con regiones.
One of the most important problems that arises when specifying individual pixels or CCD regions to be readout with different number of skipper samples $N$ is that the baseline of the pixel fluctuates, in principle, in an unpredictable way. Baseline could be considered as a swing of the mean of the pixel value $P_i$ (even when pixels are empty, with no charge). This changes in the mean of the pixels are induced by clocks and the associated circuitry and by the AC coupling of the video signal to the readout electronics. Since the CCD, as it names indicates, is a highly capacitively coupled device, the video signal is in transient response to all the clocks stimulus. In an smart readout, the clocks sequences can change for every pixel or specified ROI and because of these the baseline is constantly changing. 
Figure \ref{fig:VideoSamples_BL} shows this effect on one pixel of the video signal. Since the dynamics involved are usually slow compared to the pixel time, the net effect can be simplified as if pixel signal is mounted over a straight line (compare to the pixel in Fig.~\ref{fig:VideoSamples}). Both pixels are empty, but the one shown in Fig.~\ref{fig:VideoSamples_BL} has a trend that result in a nonzero baseline value, $P_i\approx BL_{i}$ (were $BL_{i}$ is the base line value of pixel $P_i$), when the pixel is computed with (\ref{eq:Pi}), as suggested by the pedestal and signal averages drawn with dashed lines in the figure.

\begin{figure}
    \centering
    \includegraphics[width=0.5\textwidth]{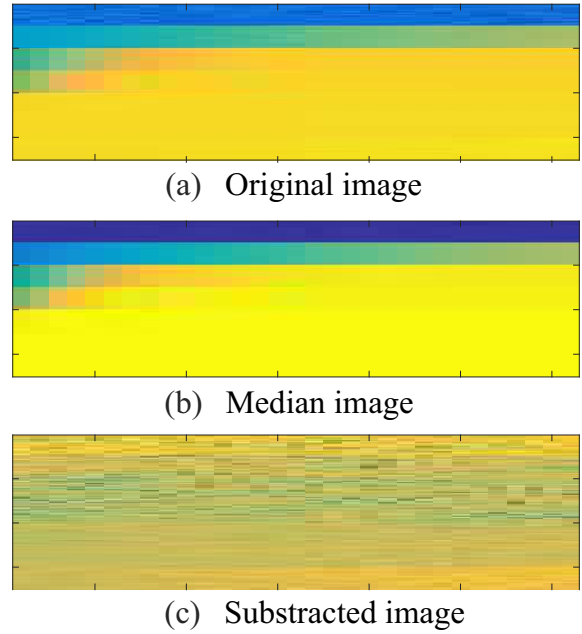}
    \caption{Illustrative process for removing the baseline. Noise-only image obtained with the Skipper-CCD with simple rectangular regions of interest. a) original image of size $31\times 175$ pixels measured with seven different regions of size $25 \times 31$ each with $N=1,10,50,100,500,1000,5000$. b) Median image obtained with several images like the one shown in a). c) Final image obtained by subtracting the median image to the original image.}
    \label{fig:ZonasRectangulares}
\end{figure}

To illustrate this, Fig.~\ref{fig:ZonasRectangulares}(a) shows $175\times 31$ pixels (rows $\times$ columns) where $7$ different rectangular ROIs of size $25 \times 31$ with $N=1,10,50,100,500,1000,5000$ were measured. Since the noise in each of these regions is different a slightly different texture is expected; however, the regions observed at first glance are mostly due to baseline variations.

\begin{figure}
    \centering
    \includegraphics[width=0.95\textwidth]{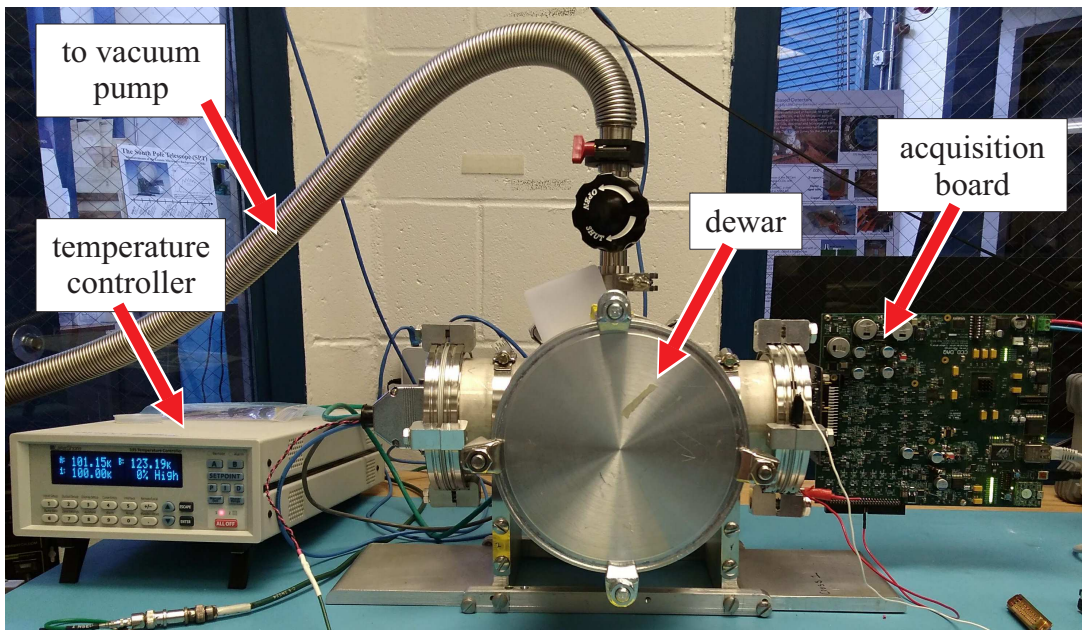}
    \caption{Photograph of the experimental setup indicating some of the main components.}
    \label{fig:Setup}
\end{figure}

The baseline should be removed to generate the final image that has different regions of interest. One way to do this is by acquiring several empty images (no charge collected) with the same regions of interest, to estimate the baseline of the pixel. Since the baseline could be considered a time invariant process, it can be estimated by computing the statistical median of each pixel for all the images. If $Q$ images are acquired, each pixel value $P_{i,skp}$  has $Q$ values one per image acquired (not to be confused with the skipper samples $N$): $P_{i,skp}^1,P_{i,skp}^2,\cdots,P_{i,skp}^Q$ the baseline of the $i$-th pixel is 
\begin{equation}
    BL_{i}\approx \mbox{median}(\{P_{i,skp}^1,P_{i,skp}^2,\cdots,P_{i,skp}^Q\}).
\end{equation}
The median is used instead of the mean to reduce the influence of outliers in the estimation, such as possible charge generated in some of the pixels.

\begin{figure*}
    \centering
    \includegraphics[width=0.95\textwidth]{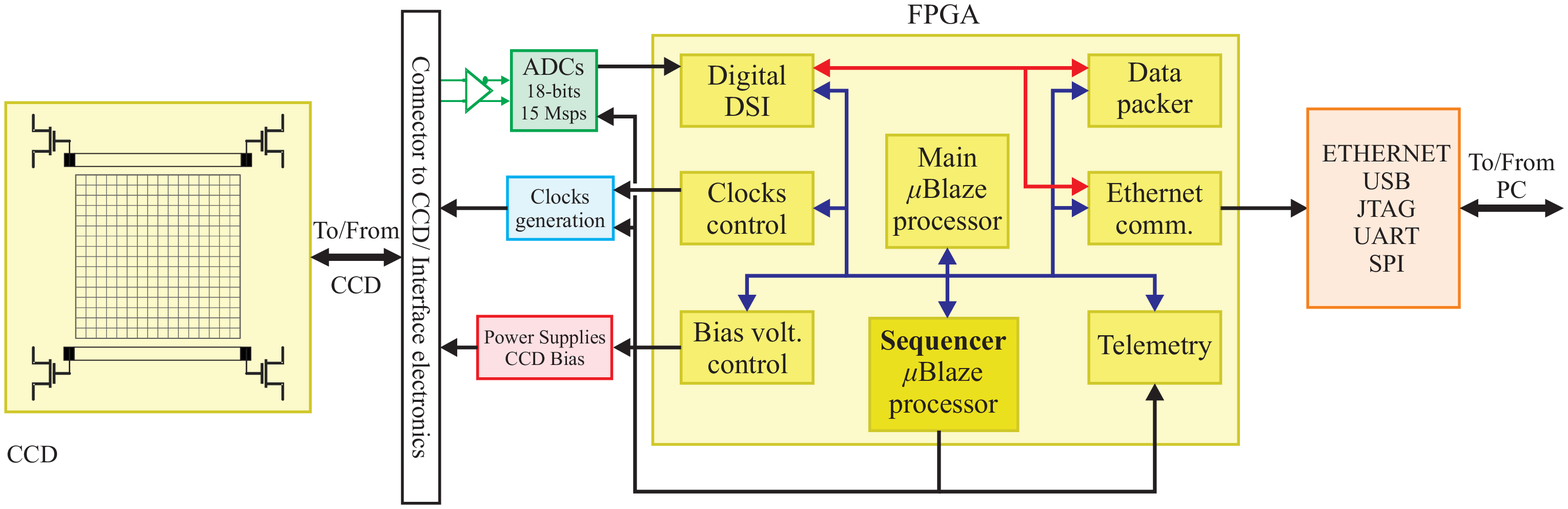}
    \caption{General diagram of the readout electronics used to implement the smart readout system with ROI. The main block is the FPGA which among other firmware block has $\mu$Blaze processor to implement the sequencer for smart readout.}
    \label{fig:DiagramaGeneral}
\end{figure*}

Figure \ref{fig:ZonasRectangulares}b shows an example of a median image where $BL_{i}$ was estimated for every pixel and Figure \ref{fig:ZonasRectangulares}c presents the final image where the baseline image was subtracted.

%\section{BASLINE CORRECTION METHOD}
%Metodo post procesamiento: Mediana

%Metodo en tiempo real: metodo de la recta. Poner ecuaciones de ruido. Ver el caso de 2 regiones de pedestal y 1 sola de señal (presentarlo aunque no lo usemos).

%ventajas y desventajas y perspectiva futura de smart dinamico.

\section{Experimental results}
\label{sec:Exp}

\begin{figure}
    \centering
    \subfloat[]{\includegraphics[width=0.37\textwidth,height=0.36\textwidth]{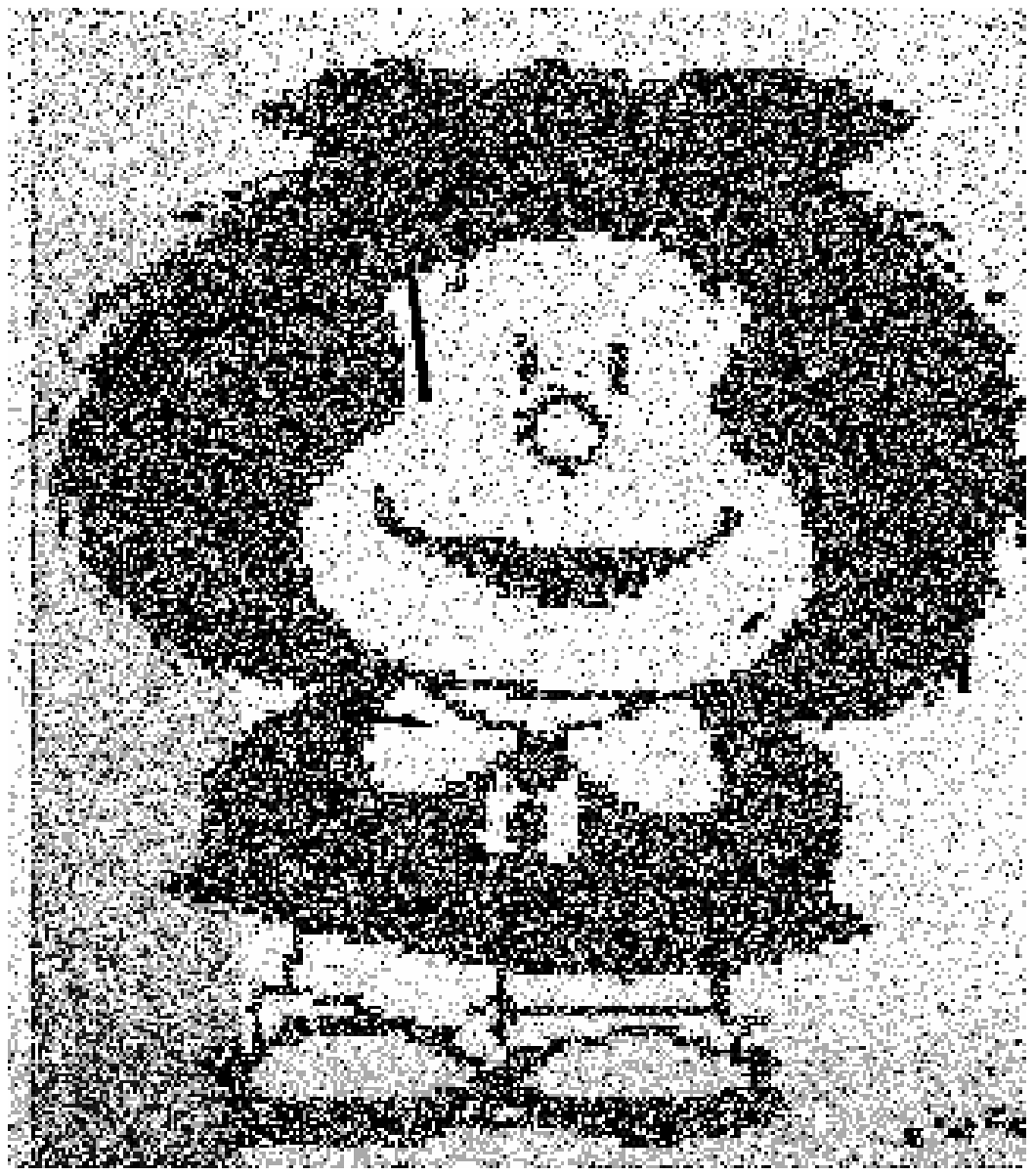} \label{Fig:Mafalda}}
    \hfill
    \subfloat[]{\includegraphics[width=0.37\textwidth,height=0.36\textwidth]{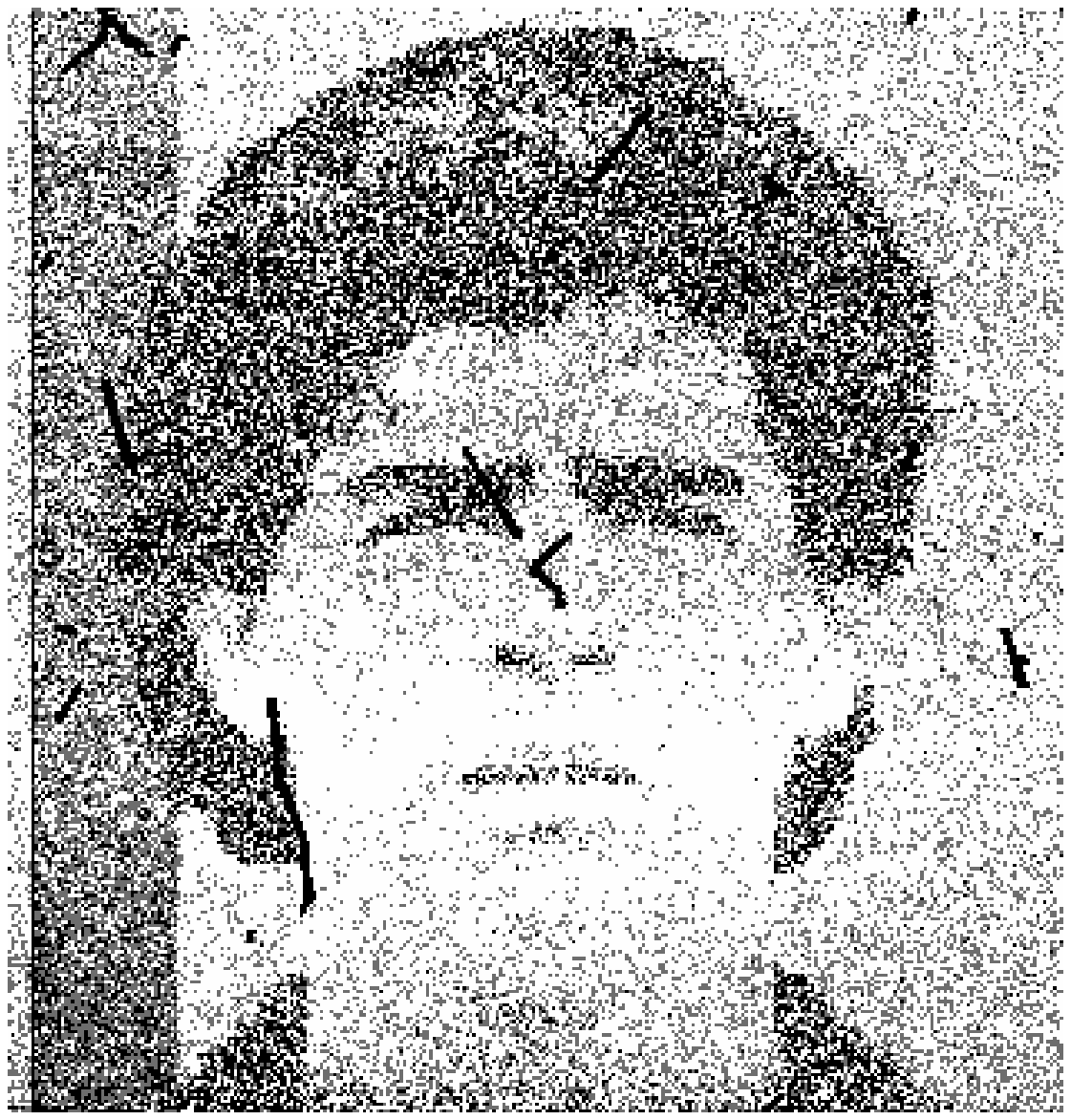}
    \label{Fig:diego}}
    \caption{Noise pictures measured with the skipper CCD with the number of samples per pixel $N$ following the patterns given from: \ref{Fig:Mafalda} a cartoon character; \ref{Fig:diego} the outline of a man. The textures that make possible to recognize the characters are caused by the different readout noise in different regions of the CCD. The image were taken in the active region of the skipper CCD and some particle interactions (electrons and muons) are also observed. }
    \label{Fig:SmartImages}
\end{figure}

\begin{figure}[t]
    \centering
    \includegraphics[width=0.95\textwidth]{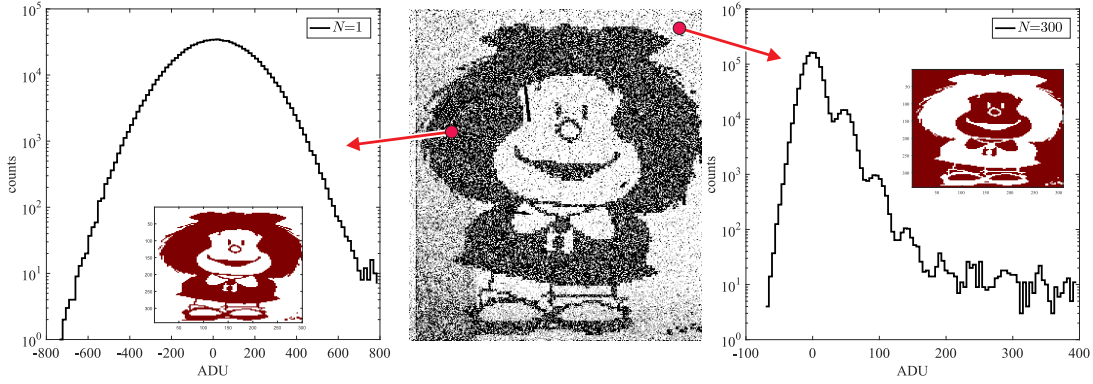}
    \caption{Histogram for the measurement of the cartoon character picture in the two different $N=1,300$ regions. For $N=1$ no charge quantization can be observed but for $N=300$ the charge peaks can be appreciated. The inset in the histograms shows which pixels were measured with the correspondent $N$ value.}
    \label{fig:MafaldaHist}
\end{figure}

\begin{figure*}
    \centering
    \includegraphics[width=0.95\textwidth]{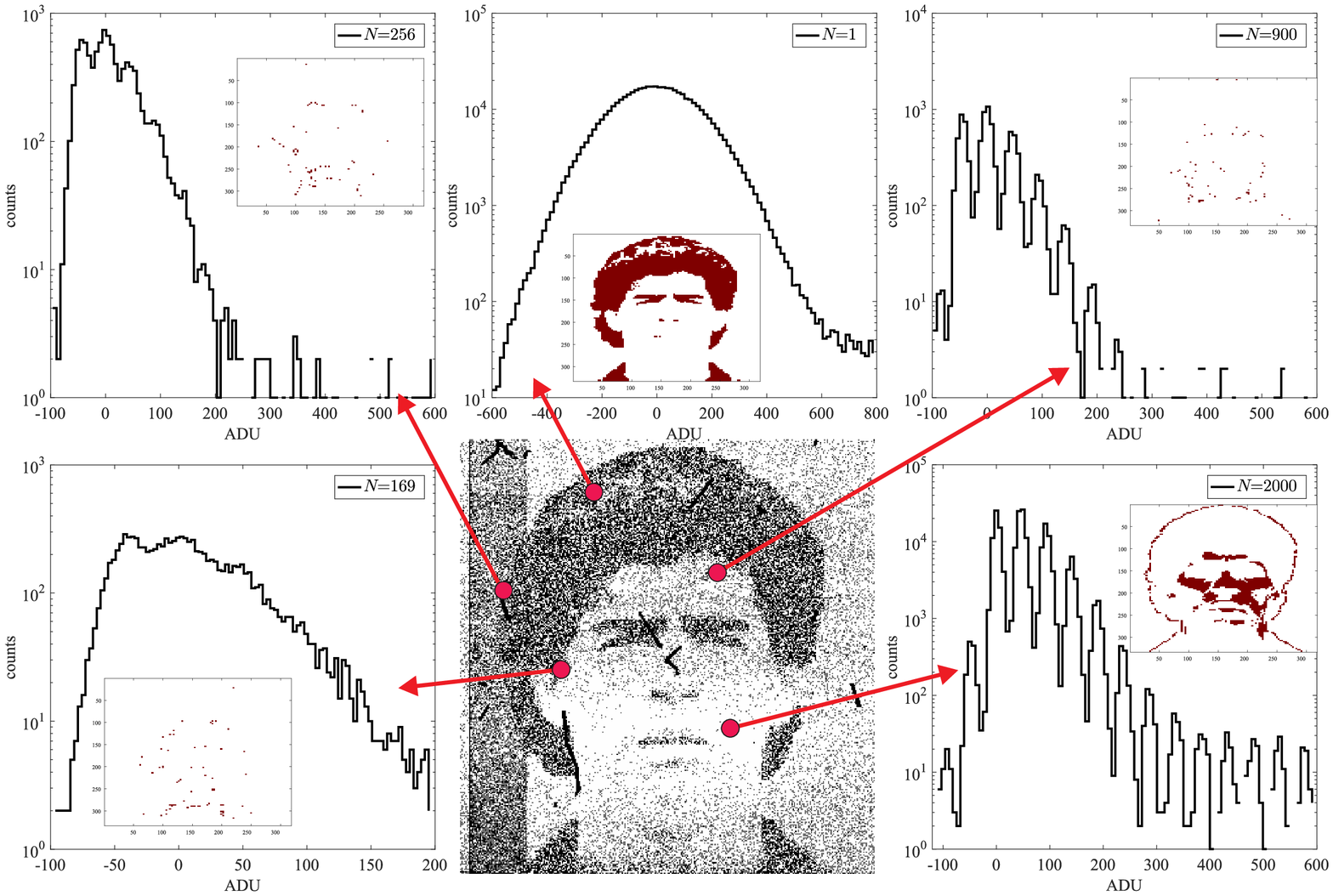}
    \caption{Histogram for the measured image of the outline of a man, where $41$ different $N$ values were used. The image shows histograms for some of them. The inset shows which pixels were measured with the correspondent $N$ value.}
    \label{fig:DiegoHIST}
\end{figure*}

In this section we present the experimental results obtained with the smart skipper readout using regions of interest with different values of skipper samples $N$. Figure \ref{fig:Setup} show a photograph of the experimental setup where the most important equipment have been labeled.  Inside the dewar, the skipper-CCD is working at high vacuum (approximately $ 10^{-4}$ mbar) and at a temperature of  around $110$K. The sensor is a fully-depleted CCD (technique developed by Lawrence Berkeley National Laboratory, LBNL) with a Skipper readout output stage (designed by LBNL).
The low threshold acquisition controller was developed by the authors \cite{cancelo2020low}, firmware modifications were specially made to implement the smart skipper readout electronics. 
Figure \ref{fig:DiagramaGeneral} depicts a block diagram of the electronics. The data processing and control of the peripheral components is performed by the Artix-7 FPGA. Bias voltage together with clock generation units create the signals that are necessary to drive the CCD, which has four output video channels. These video signals are digitized using 18-bit, 15~MSPS analog-to-digital converter based on successive approximation registers (SAR), tightly coupled with low-noise differential operational amplifiers. Output samples of the converters are fed into the FPGA to perform digital dual slope integrator (DSI) as in (\ref{eq:Pi}) to compute the pixels. The user interacts with the board through a single Ethernet port, which allows sending and receiving commands as well as data.
There are two $\mu$Blaze processor instantiated in the FPGA: one is the central unit in charge of the control of the system a peripherals and the other is in charge of the sequencer and smart skipper readout. The users specifies the sequencer including the $N$ value for each region or pixel using a pseudo-XML language description that includes environments (called recipes) to facilitate the specifications of groups of sequences which are repeatedly executed.

To show the versatility of the smart skipper readout, Fig.~\ref{Fig:SmartImages} shows to sample images acquired using the  region of interest method for smart readout. In these images the values of $N$ was specified for sub-regions of $3\times 3$ pixels and assigned based on the patterns given from: Fig.~\ref{Fig:Mafalda} a cartoon character and Fig.~\ref{Fig:diego} the outline of a man. The textures that make possible to recognize the characters are caused by the different readout noise in different regions of the CCD. The image were taken in the active region of the skipper CCD and some electrons and muons particle interactions (curvy and stight lines) are also observed. In the case of Fig.~\ref{Fig:Mafalda} $N=1$ and $N=300$ were used while for Fig.~\ref{Fig:diego} more than forty different values were used:
\begin{align}
   N=&1,4,9,16,25,36,49,64,80,81,100,121,144,169,196, \nonumber \\
     &225,256,289,324,361,400,441,484,529,576,625,676, \nonumber \\
      &729,784,841,900,961,1024,1156,1225,1369,1444,\nonumber \\
      &1600,1764,1936,2000.   \nonumber 
\end{align}
Despite the visual impact of the smart skipper readout shown in Fig.~\ref{Fig:SmartImages} which reveals how noise is different in the different regions, the histograms for pixels with same $N$ value were computed in both images.

Forty images were measured for each of the two pictures. Half of images were used to estimate the baseline as specified in \ref{Sec:Baseline} and the other half to compute the statistics of the pixels. To do so, pixels measure with the same $N$ value in the different image were grouped to compute the histograms.

For the picture of the cartoon character the results are shown in Fig.~\ref{fig:MafaldaHist}: the specified $N$ value is shown in the legends of the plot, and the correspondent pixels are shown in the inset images in the histograms. The units of the $x$-axis are in analog to digital converter units (ADUs). For $N=1$ no peaks are observed because the noise is higher than the charge of the electron. However, for $N=300$ several peaks can be observed in the histogram, those peaks correspond to $0$, $1$, $2...$, electrons of charge. 

For the picture of the outline of a man, the results are shown in Fig.~\ref{fig:DiegoHIST} for some representative values of $N$ out of the $41$ different values that were used for the measurement. As for the case of the cartoon character regions for $N=1$ do not show charge quantization but as $N$ grows charge quantization results more evident.
These results shows the capabilities of the smart skipper readout for specifying  regions of interest and perform measurements with different precision in each one.

\begin{figure}
    \centering
    \includegraphics[width=0.95\textwidth]{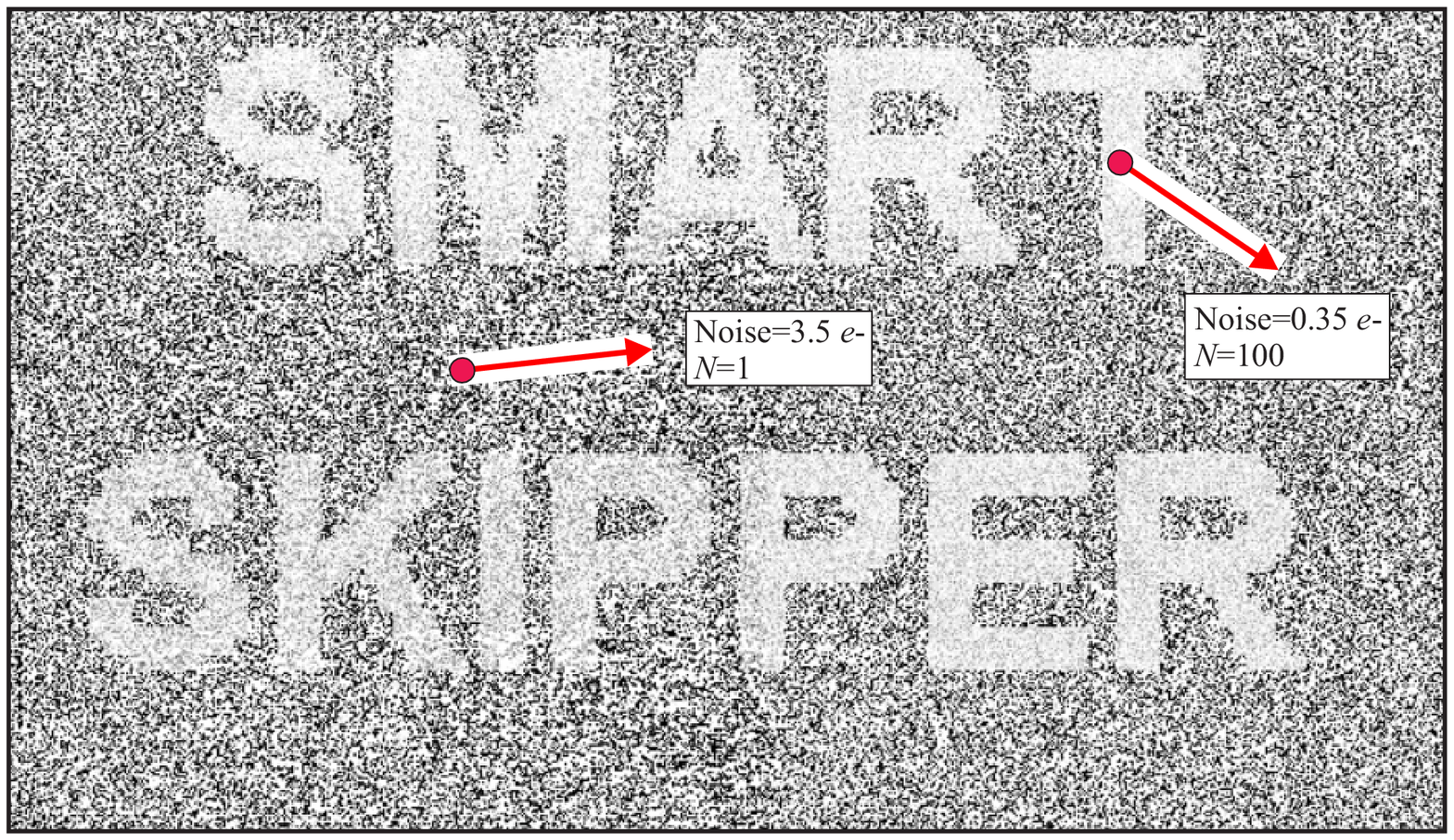}
    \caption{Noise image acquired with the smart skipper technique. The regions where ``SMART SKIPPER'' can be readout have $N=100$ resulting in a noise of $0.35e^{-}$. The rest of the pixel were measured with $N=1$ and a noise of $3.5e^{-}$. The ratio of noise reduction is $\sqrt{100}=10$ as expected from (\ref{Ec:ReduccionRuido}).}
    \label{fig:SmartSkipper}
\end{figure}

Finally Fig.~\ref{fig:SmartSkipper} shows an image acquired with $N=1$ (noise $3.5e^{-}$) and $N=100$  (noise $0.35e^{-}$) with pixels chosen to write the words ``SMART SKIPPER''. The noise in the $N=100$ region is, as expected, related to the noise of the region with $N=1$ as $3.5/\sqrt{100} e^{-}$.

\section{Conclusions}
\label{sec:conclu}
The experimental results of a smart readout technique of a skipper CCD in which the number of independent measurements of the charge of each pixel (and therefore the readout noise) can be specified based on region of interest was presented.
The development has applications in fields were the total readout time should be reduced and there are specific regions of interest in the skipper-CCD such as in astronomy, quantum imaging, etc.

\section*{Acknowledgment}
The CCD development work was supported in part by the Director, Office of Science, of the U.S. Department of Energy under Contract No. DE-AC02-05CH11231.

% References
\bibliography{main} % bibliography data in report.bib
\bibliographystyle{plain} % makes bibtex use 

\end{document}